\documentstyle[12pt]{article}
\begin{document}
\def\strut{\rule[-.5cm]{0cm}{1cm}}
\def\dspace{\baselineskip = .29in}

\title{Leptonic Fritzsch Matrices and\\
Neutrino Oscillations\thanks
{Supported in
part by Department of Energy Grant \#DE-FG02-91ER406267}}

\author{{\bf K.S. Babu}\\
\\and\\\\
{\bf Q. Shafi}\\
Bartol Research Institute\\
University of Delaware\\
Newark, DE 19716, USA}

\date{ }
\maketitle

\begin{abstract}
It was recently shown by the authors that the Fritzsch ansatz for
the quark mass
matrices prescribed at the supersymmetric grand unified scale
is compatible with a moderately heavy top quark ($m_t \simeq
120-150~GeV$).  Here we
extend the ansatz to incorporate the charged leptons and the neutrinos.
It is found that the $\nu_e-\nu_\mu$ mixing angle is small and
consistent with the
MSW solution of the solar neutrino puzzle.  Furthermore, the model
predicts  observable $\nu_\mu-\nu_\tau$ oscillations with
sin$^22\theta_{\mu \tau} \simeq 0.1$ and
$\nu_\tau$ mass in the $(1-3)~eV$ range.

\end{abstract}
\newpage

\dspace

Employing symmetries to constrain the form of
quark and lepton mass matrices
is an attractive concept, since it often leads to
relations involving the otherwise arbitrary fermionic observables
(masses, mixing angles and CP--violating phases).  A particularly simple
and elegant texture for the quark mass matrices was
proposed some time ago for three families of fermions by Fritzsch.$^1$
Only the heaviest (top) family  has a direct mass in this
scheme, while the lighter family
masses are generated via nearest neighbor mixing.  Generalizing to
include the charged leptons, the mass
matrices $M_{u,d,\ell}$ for the up--quarks, down--quarks and
charged--leptons are given by
\begin{eqnarray}
M_{u,d,\ell} = P_{u,d,\ell}\left(\matrix{0 & a_{u,d,\ell} & 0
\cr a_{u,d,\ell} & 0 & b_{u,d,\ell} \cr 0 & b_{u,d,\ell} &
c_{u,d,\ell}}\right)Q_{u,d,\ell}~~.
\end{eqnarray}
Here $P_{u,d,\ell}$ and $Q_{u,d,\ell}$ are diagonal phase
matrices and $a,b,c$ are real (positive) quantities.  The zero's of
these matrices are enforced by certain
symmetries (either discrete or
continuous).  Parity invariance is used to ensure the symmetrical nature
of the magnitudes of the various elements.  Models which
generate such textures include those with left--right symmetry,$^{1,2}$
as well as
$SO(10)$ where parity is a spontaneously broken symmetry.$^3$

In the quark sector, the Fritzsch ansatz leads to predictions for the
Cabibbo-Kobayashi-Maskawa (CKM) quark mixing angles
in terms of the quark mass ratios and
two phase parameters (denoted by $\sigma$ and $\tau$).
Even though the original mass matrices have several more phase
degrees of freedom, only the two
are observable.  All others can be rotated away by redefining the
fermion fields.  Of special concern is the prediction of the Fritzsch
ansatz for
$|V_{cb}|$, which takes the form
\begin{equation}
|V_{cb}^0| = \left| \sqrt{{{m_s^0}\over {m_b^0}}}- e^{i (\sigma-\tau)}
\sqrt{{{m_c^0}\over {m_t^0}}} \right|~~.
\end{equation}
Here the superscript on the quantities
is used to emphasize the fact that the relation holds at
whatever scale the Fritzsch texture holds.  A natural scale
at which the symmetries of the Fritzsch ansatz may be
broken is the supersymmetric grand unification (SUSY GUT)
 scale $M_G \simeq 10^{16}
{}~GeV$.  Implications of eq. (2) at low energies are then to be
evaluated by evolving the quark mass ratios and the mixing angles from $M_G$ to
the weak scale, using the renormalization group equations.

If eq. (2) holds at the weak scale, it can be verified that
the experimental value of
$|V_{cb}| = (0.043 \pm 0.009)$
sets an upper limit of about 90 GeV on the
top quark mass.$^4$  This comes about since the first term on the
right--hand side in eq. (2)
is at
least 0.15, and so needs a large
cancellation from the second term to agree with the observed value of
$|V_{cb}|$.
Such a low value of $m_t$ is on the verge of being excluded
by the
CDF search for the top quark, which sets a lower limit on $m_t$ of
91 GeV.$^5$  It also is in conflict with estimates from the one--loop
radiative corrections within the standard model, which
prefers a moderately heavy top, say in the range of $(120-160)~ GeV$.$^6
$
Could one therefore conclude that the ansatz of eq. (1) has been excluded by
experiments?

In a recent paper$^7$ we have shown that if the Fritzsch ansatz is
prescribed at a supersymmetric grand unified scale $M_G \simeq
10^{16}~GeV$, relation (2) can in fact lead to a low energy prediction
for $|V_{cb}|$ which is
consistent with observations even for a moderately heavy top quark
($m_t \stackrel{_<}{_\sim} 150~GeV$).
(Throughout this paper we shall denote by $m_t$ the
running mass
$m_t(m_t)$, which is related to the pole mass by $m_t^{\rm pole} =
m_t(m_t)[1+{{4 \alpha_s(m_t)}\over {(3 \pi)}}]$).
For large values of tan$\beta$ (the ratio of the
vacuum expectation values of the two higgs doublets which SUSY mandates)
relation (2) renormalizes in a desirable way making it consistent with
observations.  For example,
the renormalized relation for $|V_{cb}|$
at the weak scale reads (for $m_t = 140~ GeV$ and tan$\beta = 60$)
\begin{equation}
|V_{cb}| = \left| 0.89 \sqrt{{{m_s}\over {m_b}}} - 1.12e^{i (\sigma - \tau)}
\sqrt{{{m_c}\over {m_t}}}\right|~~.
\end{equation}
(Quantities without the superscripts refer to their weak scale
values.)
With an optimal choice$^8$ of $m_s(1~GeV) = 120~MeV$,
$m_b(m_b) = 4.35~GeV,~m_c(m_c) =
1.32~GeV$ and $\sigma=\tau$,
we find that $|V_{cb}| = 0.046$, in agreement with
observations.  (Without the renormalization factors, the value for $|V_{cb}|$
corresponding to the same input numbers would have been $\simeq
0.073$!)

Another testable prediction of the Fritzsch ansatz for quarks is for the
ratio$^7$ $|V_{ub}|/|V_{cb}| \simeq \sqrt{m_u/m_c} \simeq 0.06$ which is
in good agreement with the recent CLEO-II data on charmless $B$--meson decay.
The relation for the Cabibbo angle, $|V_{us}| \simeq \left|\sqrt{m_d/m_s}
- e^{i \sigma} \sqrt{m_u/m_c}\right|$, fixes the phase $\sigma$ to be
$\simeq \pi/2$.

The purpose of this paper is to extend the Fritzsch ansatz to the charged
lepton and neutrino sectors.  We are motivated by the success of the
ansatz in the quark sector
 when prescribed within a SUSY GUT framework: not only does it predict
an acceptable value for $|V_{cb}|$, in the process it also fixes the
important
parameter tan$\beta$ to be large, close to 60.
(This value
corresponds to the infra--red fixed point solution for the $b$-quark
Yukawa coupling $h_b$.$^7$)  An extension of the
ansatz to the lepton sector can
lead to predictions for the down--quark masses in terms of the charged
leptons as well as determine some of the neutrino oscillation
parameters.  The $\nu_\mu-\nu_\tau$ system turns out to be a
particularly interesting case.

We envision the matrices of eq. (1) as arising from an underlying SUSY
$SO(10)$ type grand unification.  It is well known that the
quark--lepton
symmetry$^9$ embedded in $SO(10)$ enables one to relate the quark masses
with those of leptons.  For the charged leptons, we will assume that
the mass matrix has the Fritzsch form as in eq. (1), with the (33), (12) and
(21) elements arising from a {\bf 10}--plet of Higgs.
Thus we identify
\begin{equation}
a_\ell = a_d~,~~~~c_\ell = c_d~~.
\end{equation}
The (23) and (32) elements, on the other hand, should receive
contributions from a {\bf 10} as well as a {\bf 126} of Higgs.  If only
a {\bf 10} contributed to the (23) and (32) entries, one would have the
asymptotic relation $m_s^0 = m_\mu^0$, which,
after the renormalization group corrections, is phenomenologically
unacceptable.
Similarly,
if only a {\bf 126} contributed to the (23) and (32) elements,
$m_\mu^0 = 9 m_s^0$ will follow, which again is unacceptable.

Relations (4) lead to two successful asymptotic predictions given by
\begin{equation}
m_b^0-m_s^0+m_d^0 = m_\tau^0-m_\mu^0+m_e^0;~~m_d^0m_s^0m_b^0 =
m_e^0m_\mu^0m_\tau^0~~.
\end{equation}
For
$m_t = 130~GeV$ and tan$\beta=60$,
the first of these relations predicts $m_b(m_b) \simeq 4.2~GeV$
which is in good agreement with
the spectroscopic determinations.$^8$  The second relation  leads to
$m_d(1~GeV) \simeq 7~MeV$ (if $m_s(1~GeV) = 140~MeV$), also in good
agreement with observations.  Note that the relations in
eq. (5) are identical to two of the
predictions of the Georgi--Jarlskog ansatz.$^{10}$

Before discussing the neutrino sector, one remark is in order regarding
the phase matrices $P_\ell,~Q_\ell$ and their relationship with the
matrices $P_d$ and $Q_d$ of eq. (1).  Although these phases were not
relevant for the determination of the mass eigenvalues, they do play a
role in the mixing angles.  Without loss of generality, we
can choose $P_\ell = P_d$ and
$Q_\ell=Q_d$,
provided that we allow the parameter $b_l$ to be
complex, $b_\ell=|b_\ell|e^{i \alpha}$, which is what we do in the
following.

Turning next to the neutrino sector, we assume the Dirac neutrino matrix
$M_\nu^D$ to have the Fritzsch form.  We shall further assume that
the elements of $M_\nu^D$
arise from a Higgs {\bf 10}--plet, resulting in the identity
\begin{equation}
M_\nu^D = M_u~~.
\end{equation}
The light neutrino masses depend on $M_\nu^D$ as well as on the form of the
heavy Majorana matrix $M_\nu^M$ for the right--handed neutrinos
($\nu_R$'s).  To arrive at a simple and
predictive spectrum, a judicious choice of $M_\nu^M$ is needed.  (See
refs. 11--13 for specific examples.)
We first note that a {\bf 126}--plet of Higgs
which generates such mass entries, was already used in the (23) and (32)
elements of $M_d$ and $M_\ell$.  The simplest possibility is then to use
the same {\bf 126} to generate (23) and (32) entries in the $\nu_R$
Majorana matrix.  Now $M_\nu^M$ should have rank three in order to make the
see--saw mechanism effective for all three neutrino species.  Keeping the
number of
parameters at a minimum, this is best done by allowing a non--zero
(11) element in
$M_\nu^M$.  So our extension of the Fritzsch ansatz to the heavy
Majorana matrix is given by
\begin{eqnarray}
M_\nu^M = \left(\matrix{M^{\prime} & 0 & 0 \cr 0 & 0 & M \cr
0 & M & 0}\right)~~.
\end{eqnarray}
The elements $M,~M^{\prime}$ are in general complex.

We are now in a position to discuss neutrino oscillations within the
model.  Let us first absorb the phase matrices $P_{u,d},~Q_{u,d}$
from $M_u,~M_d,~M_\ell$ and $M_\nu^D$.  Since $P_u \ne P_d$, this would
alter the charged current matrix from an identity to the matrix
\begin{eqnarray}
\left(\matrix{1 & 0 & 0 \cr 0 & e^{i \sigma} & 0 \cr 0 & 0 & e^{i \tau}}
\right)
\end{eqnarray}
both in the quark and in the leptonic charged
currents.  Note that this phase rotation would alter the phases of
$M$ and $M^{\prime}$ in eq. (7), but they were anyway arbitrary to begin with.
The advantage of this way of proceeding is that some of the phase parameters in
the leptonic CKM matrix will be the same as those in the quark sector.
If we now make a further redefinition of lepton fields so that $M_l$ becomes
real (i.e., remove the phase $\alpha$ from $|b_\ell|e^{i \alpha}$),
the leptonic charged current will have the form
\begin{eqnarray}
K = \left(\matrix{ e^{i \alpha} & 0 & 0 \cr 0 & e^{i(\sigma-\alpha)} & 0 \cr
0 & 0 & e^{i \tau} }\right)~~.
\end{eqnarray}

The light neutrino matrix $M_\nu^{light}$ is obtained from the see--saw
formula as
\begin{equation}
M_\nu^{light} = M_\nu^D (M_\nu^M)^{-1}(M_\nu^D)^T~~.
\end{equation}
Making use of the relations
\begin{equation}
c_u \simeq m_t^0,~~b_u \simeq (m_c^0m_t^0)^\frac{1}{2},~~a_u \simeq
(m_u^0m_c^0)^\frac{1}{2}
\end{equation}
(similar relations hold for other sectors), $M_\nu^{light}$ can be expressed as
\begin{eqnarray}
M_\nu^{light} \simeq {{({m_t^0})^2}\over M}
\left(\matrix{0 & \sqrt{{{m_u^0} \over {m_c^0}}}{{m_c^0}
\over {m_t^0}} & \sqrt{{{m_u^0}\over {m_t^0}}}\sqrt{{{m_c^0} \over
{m_t^0}}}
 \cr \sqrt{{{m_u^0}\over {m_t^0}}} {{m_c^0}\over {m_t^0}} &
r e^{i \gamma} {{m_u^0}\over {m_t^0}} {{m_c^0}\over {m_t^0}} &
{{m_c^0}\over {m_t^0}} \cr \sqrt{{{m_u^0}\over {m_t^0}}}\sqrt{{{m_c^0} \over
{m_t^0}}} & {{m_c^0}\over {m_t^0}} & 2\sqrt{{{m_c^0}\over {m_t^0}}}}
\right)~~.
\end{eqnarray}
Here $r=|M^{\prime}/M|$ and $\gamma$ is their relative phase.

The eigenvalues of the light neutrino mass matrix are readily obtained:
\begin{eqnarray}
m_{\nu_1} & \simeq & {{(m_t^0)^2}\over M}\left({{m_u^0} \over
{m_t^0}}\right)^2r \nonumber \\
m_{\nu_2}  & \simeq & {1 \over 2} {{(m_t^0)^2}\over M}
\left({{m_c^0}\over {m_t^0}}\right)^\frac{3}{2} \nonumber \\
m_{\nu_3} & \simeq & 2{{(m_t^0)^2}\over M}\sqrt{{{m_c^0}\over
{m_t^0}}}~~.
\end{eqnarray}

Note that in eq. (12) the parameter $r$ is accompanied  with a very small
coefficient $(m_u^0/m_t^0)(m_c^0/m_t^0)$.  It becomes relevant only
in determining the mass of
$\nu_1$ which, given the
hierarchy in the masses,
is unimportant for neutrino oscillations .  Similarly, the phase $\gamma$ is an
irrelevant variable, disappearing from all physical observables.

The leptonic CKM matrix is obtained from
\begin{equation}
V_{KM}^{lepton} = O_\nu^TKO_\ell
\end{equation}
where $K$ is the phase matrix of eq. (9), and $O_\nu$ and $O_\ell$ are
the orthogonal matrices that diagonalize $M_\nu^{light}$ and $M_l$:
$O_\ell^T M_\ell O_\ell = M_\ell(diagonal)$ and $O_\nu^T M_\nu^{light}
O_\nu = M_\nu^{light}(diagonal)$.
They are given by
\begin{eqnarray}
O_\nu \simeq \left(\matrix{1 & -\sqrt{{{m_u^0}\over {m_c^0}}} &
\frac{1}{2} \sqrt{{{m_u^0}\over {m_t^0}}} \cr
\sqrt{{{m_u^0}\over {m_c^0}}} & 1 & \frac{1}{2}\sqrt{{{m_c^0}\over
{m_t^0}}} \cr -\sqrt{{{m_u^0}\over {m_t^0}}} & -\frac{1}{2}
\sqrt{{{m_c^0}\over{m_t^0}}} & 1}\right)
\end{eqnarray}
\begin{eqnarray}
O_\ell \simeq \left(\matrix{1 & -\sqrt{{{m_e^0}\over {m_\mu^0}}} &
\sqrt{{{m_e^0}\over {m_\tau^0}}}{{m_\mu^0}\over {m_\tau^0}} \cr
\sqrt{{{m_e^0}\over {m_\mu^0}}} & 1 & \sqrt{{{m_\mu^0}\over
{m_\tau^0}}} \cr -\sqrt{{{m_e^0}\over {m_\tau^0}}} &
-\sqrt{{{m_\mu^0}\over {m_\tau^0}}} & 1}\right)~.
\end{eqnarray}
The resulting leptonic CKM elements are
\begin{eqnarray}
|V_{1\mu}| & \simeq & |V_{2 e}| \simeq \left|\sqrt{{{m_e^0}\over
{m_\mu^0}}}-e^{i(\sigma-2 \alpha)} \sqrt{{{m_u^0}\over {m_c^0}}}
\right| \nonumber \\
|V_{2 \tau}| & \simeq & |V_{3 \mu}| \simeq \left|\sqrt{{{m_\mu^0} \over
{m_\tau^0}}} - \frac{1}{2}e^{i(\tau-\sigma+\alpha)}\sqrt{{{
m_c^0}\over {m_t^0}}}\right| \nonumber \\
|V_{1 \tau}| & \simeq & \left|\sqrt{{{m_e^0}\over {m_\tau^0}}}
{{m_\mu^0}\over {m_\tau^0}} + e^{i(\sigma - 2 \alpha)} \sqrt{{{
m_u^0}\over{m_c^0}}}\sqrt{{{m_\mu^0}\over {m_\tau^0}}}-e^{i(\tau
- \alpha)}\sqrt{{{m_u^0}\over {m_t^0}}}\right| \nonumber\\
|V_{3 e}| & \simeq & \left|\frac{1}{2}\sqrt{{{m_u^0}\over {m_t^0}}}
+\frac{1}{2}e^{i(\sigma-2\alpha)}\sqrt{{{m_c^0}\over{m_t^0}}}
\sqrt{{{m_e^0}\over {m_\mu^0}}}-e^{i(\tau-\alpha)}\sqrt{{{m_e^0}
\over{m_\tau^0}}}\right|~.
\end{eqnarray}

The leptonic mixing angles do not run below the scale of
$B-L$ breaking, since the right--handed neutrinos decouple at that
scale.  However, the relations (13) and (17)
are in terms of the asymptotic
masses of quarks and charged leptons.  It is therefore necessary to
extrapolate the low energy masses to the SUSY GUT scale using the (one
loop)
renormalization group equation for the mass ratios and mixing
angles including the effect of a heavy third family.$^{14}$
In ref. (7) a detailed analysis of this type
was carried out.  We extend
it here to include the charged lepton mass ratios.
The variations of these quantities
as
functions of the top quark mass corresponding to tan$\beta=60$
are shown in Fig. (1).
As input values we chose $\alpha_1(M_Z) = 0.01013,~\alpha_2(M_Z) =
0.03322,~\alpha_3(M_Z) = 0.115$ and $M_G = 10^{16}~GeV$.  Our numerical
results are in agreement with the analytic results presented in
ref. (15).
For $m_t = 130~GeV$ the relevant
renormalization factors are found to be
\begin{eqnarray}
\left({{m_c^0}\over {m_t^0}}\right) & = & 0.64 \left({{m_c}\over {m_t}}
\right);~~\left({{m_u^0}\over {m_t^0}
}\right) = 0.64 \left({{m_u}\over {m_t}}\right) \nonumber \\
\left({{m_\mu^0}\over {m_\tau^0}}\right) & = & 0.53 \left({{m_\mu}\over
{m_\tau}}
\right);~~
\left({{m_e^0}\over {m_\tau^0}}\right) = 0.53 \left({{m_e}\over {m_\tau}}
\right)~~.
\end{eqnarray}

{}From eq. (13), it follows that
\begin{equation}
{{m_{\nu_2}}\over {m_{\nu_3}}} = {{m_c^0}\over {4 m_t^0}}~~.
\end{equation}
Choosing $m_c(m_c) = 1.27~GeV$ and noting that the QCD running factor from
$m_c$ to $m_t$ is $\simeq 0.6$, we find that
\begin{equation}
{{m_{\nu_2}}\over {m_{\nu_3}}} \simeq 9 \times 10^{-4}~~.
\end{equation}

{}From eq. (17) one sees that the $\nu_\mu-\nu_\tau$ oscillation angle is
rather large, in the range of $(15-25)\%$.  For such a large mixing,
there is an upper limit of about 2.5 eV on the $\nu_\tau$
mass arising from oscillation experiments.$^{16}$
Eq. (20) translates this into an
upper limit on $m_{\nu_2} \le 2.3 \times 10^{-3}~eV$.  Remarkably, this
is in the right range for $\nu_e-\nu_\mu$ MSW oscillations, corresponding
to
$\Delta m^2_{\nu_e\nu_\mu} \simeq 5.4 \times 10^{-6}~eV^2$.$^{17}$
We also see that the $\nu_\tau$ mass ($m_{\nu_3}$)
cannot be less than about 1 eV,
otherwise $\nu_\mu$ would be outside of the MSW range.
A neutrino mass
in the range of $(1-3)~eV$ is cosmologically significant and
plays the role of the hot component of
dark matter, which is suggested$^{18}$ by the recent COBE data.

As for the $\nu_e-\nu_\mu$ mixing angle, first note that the ratios of
masses involving the first two families do not run.  The two terms
$\sqrt{m_e/m_\mu}$ and $\sqrt{m_u/m_c}$ are numerically about equal.
This means that the mixing angle can be small and quite
different from the naive expectation that $|V_{1 \mu}| \simeq |V_{us}|
\simeq 0.22$.  The combined SAGE/GALLEX experiments in fact prefer
a mixing angle in the $\nu_e-\nu_\mu$ sector which is
significantly smaller than $|V_{us}|$.
For a `central' value of $|V_{1 \mu}| =
0.05$ implied by SAGE/GALLEX, we can fix all the other oscillation
parameters of the model.  Using
$m_u(1~GeV) = 5.1~MeV,~m_c(m_c) = 1.27~GeV$, we determine
$\sqrt{m_u/m_c} = 0.062$, to be compared with $\sqrt{m_e/m_\mu} = 0.07$.
Then, from the first relation of eq. (17), we find cos($\sigma-\alpha$)
$ = 0.72$, or $\sigma-\alpha \simeq \pi/4$.  From the quark sector, we also
know that $\sigma \simeq \pi/2$ (to get the Cabibbo angle right),
so $\alpha \simeq \pi/4$ is preferred.
Using this value of $\alpha$ and $\tau \simeq \sigma$ (needed for
acceptable $|V_{cb}|$), we can calculate
\begin{equation}
|V_{2 \tau}| \simeq |V_{3 \mu}|
\simeq 0.158;~~|V_{3 e}| \simeq 0.011;~~|V_{1 \tau}|
\simeq 0.0098~~~.
\end{equation}
With such a significant mixing in the (2-3) sector along with the
constraint $m_{\nu_{\tau}} \ge 1~eV$, it follows
that
$\nu_\mu-\nu_\tau$ oscillations should be observable in the planned
CHORUS/NOMAD experiments at CERN and the proposed Fermilab experiment
P803.$^{19}$
The $\nu_e-\nu_\tau$ mixing may play a significant role in
astrophysical settings, for example in blowing up the supernova
core.

At this stage it is worthwhile to assess the predictive power of the
Fritzsch ansatz including the leptonic generalization proposed here and
compare its predictions with those of other popular
ansatzes.$^{20-22}$  In
the charged fermion sector there are altogether 9 parameters
($a_{u,d},~b_{u,d},~c_{u,d},~\sigma,~\tau$ and
$|b_\ell|$) and 13 observables, thereby leading to 4 predictions.  In addition,
the parameter tan$\beta$ is determined.
The
inclusion
of neutrinos adds three additional parameters, $|M|$, $|M^{\prime}|
$ (as noted earlier, their relative phase became irrelevant) and the
phase $\alpha$ associated with $b_\ell$.
There are now 9 more observables (3 neutrino masses, 3
mixing angles and 3 CP violating phases).   The 23 (= 13 + 1 + 9)
observables are determined in terms of 12 parameters, thereby resulting in
11 predictions.  In the charged fermion sector, the predictions include
the
two mass relations of eq. (5), $|V_{ub}|/|V_{cb}| \simeq \sqrt{m_u/m_c}
\simeq 0.06$, tan$\beta \simeq 60$ and the prediction for the CP
violating parameter in the CKM matrix.$^7$  In the neutrino sector,
they include the mass ratio $m_{\nu_2}/m_{\nu_3}$ of eq. (19), the two
mixing angles in eq. (21), and the three CP violating phases.  The three
distinguishing features of the Fritzsch ansatz are (i) tan$\beta$ is
large, (ii) the top quark is moderately heavy,
$m_t \simeq 120-150~GeV$,
and (iii) $|V_{cb}|$ can be close to its central
value.
At least two of the predictions should distinguish it from
the ansatz of
ref. (22), which predicts $m_t \stackrel{_>}{_\sim} 170~GeV$
and $|V_{cb}| \stackrel{_>}{_\sim} 0.052$.

In conclusion, motivated by the success the Fritzsch ansatz
enjoys in the quark sector,
we have proposed a generalization to include the leptons.$^{23}$
We found
that the minimal scheme which
incorporates the charged leptons and neutrinos predicts
a small $\nu_e-\nu_\mu$ mixing angle
which is consistent with the MSW resolution of the solar neutrino
problem.
The
$\nu_\tau$ mass is predicted to be in the range $(1-3)~eV$, making it a
suitable candidate for the hot component of dark matter.  The planned
$\nu_\mu-\nu_\tau$ oscillation experiments and the discovery of the top
quark in the mass range $m_t \simeq (120-150)~GeV$
will provide crucial tests of
the idea.

\section*{References}
\begin{enumerate}
\item H. Fritzsch, Phys. Lett. {\bf B70}, 436 (1977) and Nucl. Phys.
{\bf B155}, 189 (1979); \newline
For related earlier work see,
S. Weinberg, in Festschrift for I.I. Raby,
Transactions of the New York Academy of Sciences (1977); \newline
F. Wilczek and A. Zee, Phys. Lett. {\bf 70B}, 418 (1977).
\item See for example, K.S. Babu, Phys. Rev. {\bf D 35}, 3477 (1987).
\item See for example, G. Lazaridez and Q. Shafi, Nucl. Phys. {\bf B350},
179 (1991).
\item F. Gilman and Y. Nir, Ann. Rev. Nucl. Part. Sci. {\bf 40}, 213
(1990);
\newline E. Ma, Phys. Rev. {\bf D 43}, R2761 (1991);
\newline K. Kang, J. Flanz and E. Paschos, Z. Phys. {\bf C55}, 75
(1992).
\item F. Abe et. al., CDF Collaboration, Phys. Rev. Lett. {\bf 68}, 447
(1992).
\item J. Ellis and G. Fogli, Phys. Lett. {\bf 249}, 543 (1990);
\newline P. Langacker and M. Luo, Phys. Rev. {\bf D 44}, 817 (1991).
\item K.S. Babu and Q. Shafi, Bartol Preprint BA-92-70, Phys. Rev. {\bf D},
(to be published).
\item J. Gasser and H. Leutwyler, Phys. Rep. {\bf 87}, 77 (1982).
\item J.C. Pati and A. Salam, Phys. Rev. {\bf D 10}, 275 (1974).
\item H. Georgi and C. Jarlskog, Phys. Lett. {\bf B89}, 297 (1979);
\newline H. Georgi and D.V. Nanopoulos, Nucl. Phys. {\bf B159}, 16
(1979).
\item K.S. Babu and Q. Shafi, Phys. Lett. {\bf B294}, 235 (1992).
\item S. Dimopolous, L. Hall and S. Raby, Preprint UCB-PTH-92-21 (1992).
\item K.S. Babu and R.N. Mohapatra, Bartol Preprint BA-92-54 (1992).
\item E. Ma and S. Pakvasa, Phys. Lett. {\bf B 86}, 43 (1979) and Phys.
Rev. {\bf D 20}, 2899 (1979);
\newline K.S. Babu, Z. Phys. {\bf C 35}, 69 (1987);
\newline K. Sasaki, Z. Phys. {\bf C 32}, 149 (1986);
\newline B. Grzadkowski, M. Lindner and S. Theisen, Phys. Lett.
{\bf B198}, 64 (1987);
\newline M. Olechowski and S. Pokorski, Phys. Lett. {\bf B257}, 388
(1991);
\newline V. Barger,
M. Berger and P. Ohmann, Wisconsin Preprint MAD-PH-722 (1992).
\item S. Naculich, Johns Hopkins Preprint JHU-TIPAC-930002 (1993).
\item N. Ushida et al., Phys. Rev. Lett. {\bf 57}, 2897 (1986).
\item For recent updates, see, S. Bludman, N. Hata, D.
Kennedy and P. Langacker, Pennsylvania Preprint UPR-0516T (1992);
\newline P. Krastev and S. Petcov, Phys. Lett. {\bf B299}, 99 (1993).
\item R. Schafer and Q. Shafi, Nature {\bf 359}, 199 (1992).
\item CHORUS Collaboration, N. Armenise et. al., Preprint
CERN-SPSC/90-42 (1990);
\newline NOMAD Collaboration, P. Astier et. al., Preprint
CERN-SPSC/91-21 91991);
\newline K. Kodama et. al., Fermilab Proposal P803 (1991).
\item J. Harvey, P. Ramond and D. Reiss, Phys. Lett. {\bf B92}, 309
(1980) and Nucl. Phys. {\bf B 199}, 223 (1982);
\newline H. Arason et. al., Phys. Rev. {\bf D 46}, 3945 (1992).
\item E. Friere, G. Lazarides and Q. Shafi, Mod. Phys. Lett.
{\bf A 5}, 2453 (1990).
\item S. Dimopoulos, L. Hall and S. Raby, Phys. Rev. Lett. {\bf 68},
1984 (1992);
\newline G. Anderson, S. Raby, S. Dimopoulos and L. Hall, Preprint
OHSTPY-HEP-92-018 (1992).
\item For related discussions employing leptonic Fritzsch matrices see,
\newline A.J. Davies and X-G. He, Phys. Rev. {\bf D 46}, 3208 (1992);
\newline C.Q. Geng and J. Ng, Phys. Rev. {\bf D 39}, 1925 (1989);
\newline T. Rizzo and J. Hewett, Phys. Rev. {\bf D 33}, 1519 (1986) and
Phys. Rev. {\bf D 34}, 298 (1986);
\newline A. Joshipura, Phys. Lett. {\bf 164B}, 333 (1985).
\end{enumerate}

\section*{Figure Caption}
Fig. 1:  The running factors $f(M_X)/f(M_Z)$ for $f = |V_{cb}|$ (solid)
$|m_c/m_t|$ \\
\hspace*{.6in} (dot--dash), $|m_\mu/m_\tau|$
(solid) and $|m_s/m_b|$ (dashed) versus $m_t$.  The \\
\hspace*{.6in} running factors for $|V_{ub}|,~|V_{td}|,~|V_{ts}|$
are identical to that
of $|V_{cb}|$.\\
\hspace*{.6in} Similarly, $|m_e/m_\tau|$ runs as $|m_\mu/m_\tau|$,
$|m_d/m_b|$ runs as $|m_s/m_b|$ \\
\hspace*{.6in} and
$|m_u/m_t|$ as $|m_c/m_t|$.

\end{document}